\documentclass[twocolumn,amssymb,amsmath,floats,showpacs,superscriptaddress,prl,floatfix]{revtex4-1}

\usepackage{amsmath,amssymb,bm}
\usepackage{graphics}
\usepackage{epsfig}
\usepackage{graphicx}

\begin{document}

\title{Noise-induced phase transitions in neuronal networks}

\author{K.-E. Lee}
\affiliation{
Department of Physics $\&$ I3N, University of Aveiro,
3810-193 Aveiro, Portugal}
\author{M. A. Lopes}
\affiliation{
Department of Physics $\&$ I3N, University of Aveiro,
3810-193 Aveiro, Portugal}
\author{A. V. Goltsev}
\affiliation{
Department of Physics $\&$ I3N, University of Aveiro,
3810-193 Aveiro, Portugal}
\affiliation{Ioffe Physical-Technical Institute, 194021 St.~Petersburg, Russia}

\begin{abstract}
Using an exactly solvable cortical model of a neuronal network, we show that,
by increasing the intensity of shot noise (flow of random spikes bombarding neurons),
the network undergoes first- and second-order non-equilibrium phase transitions. We study the nature of the transitions, bursts  and avalanches of neuronal activity.
Saddle-node and supercritical Hopf bifurcations are the mechanisms of emergence of sustained network oscillations.
We show that the network stimulated by shot noise behaves similar to the Morris-Lecar
model of a biological neuron stimulated by an applied current.
\end{abstract}



\maketitle



In the brain, interactions among neurons lead to diverse collective phenomena such as phase transitions, self-organization, avalanches, and brain rhythms \cite{Kelso_1995,Chialvo_2006}.
A non-equilibrium second-order phase transition was observed in human bimanual coordination \cite{Kelso_1984,hkb1985,Kelso_1986}. Stimulation of living neural networks by electric fields causes a first-order phase transition \cite{Breskin_2006}.
There are evidences that
brain rhythms, epileptic seizures, and the ultraslow oscillations of BOLD fMRI patterns also emerge as a result of non-equilibrium phase transitions \cite{Steyn-Ross_2010}.
The Hopf and saddle-node bifurcations are generic mechanisms for the emergence of oscillations in nonlinear differential equation models \cite{Strogatz_book1994}. These mechanisms were found in the Morris-Lecar model of a biological neuron
\cite{Rinzel1989,Izhikevich2000}.
In the context of brain rhythms,
the Hopf bifurcations were discussed within mean-field cortical models \cite{Steyn-Ross_2010} and for networks of randomly connected integrate-and-fire neurons \cite{bh1999,b2000,obh2009,lb2011}. At the present time, understanding of nature of collective phenomena in the brain is elusive.
For a complete description of a phase transition it is not enough to identify the bifurcation. It is also necessary to find critical phenomena that accompany it.
In statistical physics, exactly solved models
largely help us to understand phase transitions and critical phenomena \cite{Baxter_1982}.

In the present paper, we propose
an exactly solvable cortical model with stochastic excitatory and inhibitory neurons stimulated by shot noise (a flow of random spikes bombarding neurons).
We show that shot noise stimulates first- and second-order non-equilibrium phase transitions in collective dynamics of neuronal networks. The first-order phase transition occurs as
a discontinuous  transition from low to high neuronal activity.
Avalanches precede the transition. The saddle-node and supercritical Hopf bifurcations are the mechanism of emergence of sustained network oscillations.
We find the order parameter for the continuous  phase transitions and critical phenomena that accompany them  in activity fluctuations.
We show that the model
exhibits collective excitability similar to excitability of the Morris-Lecar neuron stimulated by an applied current.

\emph{Model.} We study a cortical model
composed of $N_e$ excitatory and $N_i$ inhibitory neurons. $N_e+N_i\equiv N$ is the network size, $g_{e(i)}\equiv N_{e(i)}/N$ is the fraction of excitatory (inhibitory) neurons. Neurons are randomly connected with probability $c/N$ and form a classical random graph with Poisson degree distribution and the mean degree $c$. The network is locally tree-like and has the small-world properties \cite{Albert_2002,dg2002,newman2003} similar to ones found in brain networks \cite{sporns04}.
To take into account noise playing an important role in the brain dynamics \cite{lgns04,faisal08,ermentrout08}, we assume that
neurons are bombarded by random spikes represented by Dirac delta functions, $I(t)=\sum_{i} q \delta(t-t_i)$,
where $t_i$ are arrival times of spikes and $q$ is their amplitude. This kind of random input is a so-called shot noise. Random spikes represent spontaneous release of neurotransmitters in synapses (synaptic noise) and random spikes arriving from the remote part of the cortex.
According to Schottky's result, in the case of the Poisson distribution of
interspike intervals, the power spectral density $S(\omega)$
is proportional to the mean frequency $\omega_{sn}$ of spikes, $S(\omega)=2q^2 \omega_{sn}$.
We assume that the probability to receive $\xi$ random spikes during the integration
time $\tau$ is Gaussian, $G(\xi)=G_0 \exp[-(\xi-\langle n\rangle)^2 / 2\sigma^2]$
where $\sigma^2$ is the variance, $\langle n\rangle= \omega_{sn}\tau$ is the mean number of spikes arriving during the time interval $\tau$, and $G_0$ is the normalization constant, $\sum_{\xi=0}^{\infty}G(\xi)=1$. We will use $\langle n\rangle$ as the control parameter characterizing the shot noise intensity.

Neurons also receive delta-like
spikes from active neighbors. The spikes mediate interaction among neurons.
The total input $I(t)$ includes spikes from
shot noise and excitatory and inhibitory presynaptic neurons.
We define the input $V_n$ to a neuron with index $n$, $n=1,2,\dots N$, as integral of $I(t)$ over the time interval $[t-\tau,t]$. This gives
\begin{equation}
V_n(t)= \xi q + kJ_e +lJ_i,
\label{input}
\end{equation}
where $\xi$, $k$, and $l$ are the numbers of spikes arriving during the time interval $[t-\tau,t]$ from shot noise, active presynaptic excitatory and inhibitory neurons, respectively. We assume that efficacies of synaptic connections with excitatory and inhibitory neurons are uniform and equal to $J_e$ ($J_e >0$) and $J_i$ ($J_i <0$), respectively. The numbers $k$ and $l$ are random and are determined by activity of presynaptic neurons during the interval $[t-\tau,t]$. The network structure is encoded in the adjacency matrix.

In our model, neurons are tonic and the firing frequency $f(V)$ versus input $V$ is the Heaviside function $f(V)=f \Theta(V-V_{th})$ where $V_{th}$ is a threshold.
$f$ is the same for both excitatory and inhibitory neurons. If $f\tau <1$ and spike emission times of neurons are uncorrelated, then
during the time interval $[t-\tau,t]$, each active presynaptic neuron contributes to $V_n(t)$ either one spike with probability $\tau f$ or none with probability $1-\tau f$.

We consider stochastic neurons like those of \cite{Benayoun_2010,Wallace_2011,goltsev10,goltsev13}.
It means that response of neurons to an input is a stochastic process. Such a stochastic behavior might be caused by cellular noise and intensive bombardment by random spikes.
Two rules determines dynamics of the cortical model:
(1) If the input $V_n(t)$ at an inactive excitatory (inhibitory) neuron $n$ at time $t$  is at least a certain threshold $V_{th}$, then this
neuron is activated with probability $\mu_{e} \tau$ ($\mu_{i} \tau$) and fires spikes.
(2) An active excitatory (inhibitory) neuron $n$ is inactivated with probability $\mu_{e}\tau$ ($\mu_{i}\tau$) if $V_n(t)< V_{th}$.
In our model, $1/\mu_{e}$ and $1/\mu_{i}$ are of the order of the first-spike latencies of excitatory and inhibitory neurons (from 6 to 100 ms, in the cortex).
We introduce a parameter $\alpha \equiv \mu_{i}/\mu_{e}$.
If $\alpha <1$,
inhibitory neurons have a larger response time than excitatory neurons.
$\alpha$ may be both larger and smaller than 1.
We also introduce a dimensionless activation threshold $\Omega \equiv V_{th}/J_{e} $.
$\Omega$ is of the order of 15-30 in living neuronal networks
and about $30-400$ in the brain.

\emph{Rate equations.} Behavior of the model
is described by the fractions $\rho_e(t)$ and
$\rho_i(t)$ of active
excitatory and inhibitory neurons, respectively, at time $t$. We will call them
`activities'.
We assume that activities are changed slightly during the integration time $\tau$.
Using the rules
formulated above and the methods developed in \cite{goltsev10,goltsev13,dorogovtsev08}, in the limit $N\rightarrow\infty$, we find explicit rate equations,
\begin{eqnarray}
& \dot{\rho}_e(t)=F_{e}(t)[1{-}\rho_e(t)]-\mu_e \rho_{e}(t)+\mu_e \Psi_{e}(\rho_{e}(t),\rho_{i}(t)), \nonumber \\
& \!\!\!\! \dot{\rho}_i(t)=F_{i}(t)[1{-}\rho_i(t)]-\mu_i \rho_{i}(t) + \mu_i \Psi_{i}(\rho_{e}(t),\rho_{i}(t)),
\label{eq:10}
\end{eqnarray}
where $\dot{\rho}\equiv d \rho/dt$. $\Psi_{e(i)}(\rho_e,\rho_i)$ is the probability
that, at given activities $\rho_e$ and $\rho_i$, the input to a randomly chosen excitatory (inhibitory) neuron is at least  $\Omega$. $F_{e}$ and $F_{i}$ represent
fields acting on excitatory and inhibitory neurons. In the case of the classical random graph, we find
\begin{eqnarray}
& \Psi_i(\rho_e,\rho_i)=\Psi_e(\rho_e,\rho_i)\equiv
\Psi(\rho_e,\rho_i) = \nonumber \\
&\!\!\!\!\!\!\!\!\!\sum_{k,l,\xi=0}^{\infty}\Theta(k J_{e}{+}lJ_{i}{+}\xi q{-}\Omega J_{e}) G(\xi)
P_{k}(g_{e}\rho_{e} \widetilde{c})P_{l}(g_{i}\rho_{i} \widetilde{c}),
\label{eq:14}
\end{eqnarray}
where $\widetilde{c} \equiv c\tau f$,
$P_k(g_e\rho_e \widetilde{c})$ and $P_l(g_i\rho_i \widetilde{c})$ are the probabilities that,
during the time interval $\tau$,
a randomly chosen neuron receives $k$ spikes from excitatory and $l$ spikes from inhibitory neurons, respectively. Here $P_k(c)\equiv c^k e^{-c}/k!$.

\emph{Algorithm.} In simulations,
we builded a directed network, linking neurons with the probability $c/N$.
We divided time into intervals of width $\Delta t=\tau$. At each time step, for each neuron we calculated input Eq.~(\ref{input})
given that each active presynaptic neuron contributes a spike with probability $\tau f$. The number of random spikes (shot noise) in this input was generated
by the Gaussian process $G(\xi)$.
Then, we updated states of neurons, using the rules formulated above. In the paper, numerical calculations are presented for parameters $N=10^5$, $c=10^3$, $\Omega=30$, $\tau f = 0.1$, $f=\mu_{e}$, and $g_i=0.25$.
We use $1/\mu_{e}\equiv 1$ as time unit and $J_e\equiv 1$ as input unit. Following \cite{amit97}, we choose $J_{i}=-3J_{e}$. We use $q=J_{e}$ and $\sigma^2 =10$ for the amplitude and variance of shot noise.

\begin{figure}
\includegraphics[width=0.25\textwidth]{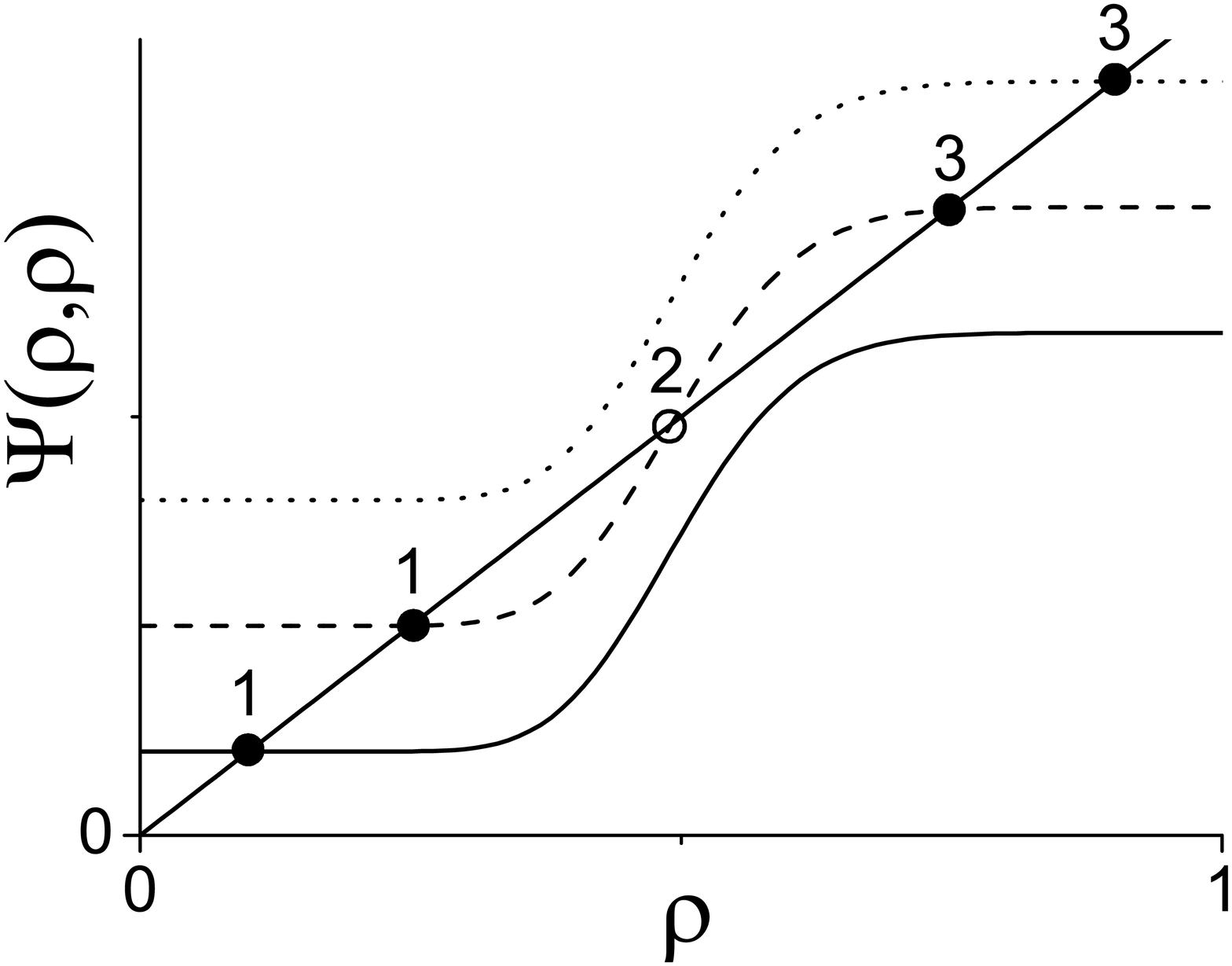}
\caption{Points 1,2, and 3 represent solutions of the steady state equation $\rho=\Psi(\rho,\rho)$ for the cases  $\langle n \rangle <n_{c1}$ (solid line),  $n_{c1} < \langle n \rangle <n_{c2}$ (dashed line), and $\langle n \rangle > n_{c2}$ (dotted line).
\label{fixpoints}}
\end{figure}

\emph{Steady states.} If $F_{e}=F_{i}=0$, then in a steady state, we have $\rho_e =\rho_i\equiv\rho$ and $\rho$ is a solution of the steady state equation, $\rho=\Psi(\rho,\rho)$.
A graphical solution of the equation is displayed in Fig. \ref{fixpoints}. If shot noise intensity $\langle n \rangle$ is either sufficiently small or sufficiently large, then there is only one solution, either point 1 or point 3. These fixed points correspond to the steady states with low and high neuronal activity, respectively. In an intermediate range $n_{c1} < \langle n \rangle <n_{c2}$ there are three fixed points (1,2, and 3). At the critical point  $\langle n \rangle = n_{c1}$, points 2 and 3 coalesce.  Points 1 and 2 coalesce at $\langle n \rangle = n_{c2}$. From Fig.~\ref{fixpoints} one sees that the coalescence occurs when $d \Psi(\rho,\rho)/d\rho=1$. Together with the steady state equation,
this condition determines $n_{c1}$ and $n_{c2}$.
While the fixed points depend on $\langle n \rangle$, but not on $\alpha$, their local stability depends on both $\langle n \rangle$ and $\alpha$. It is determined by eigenvalues of the Jacobian of Eqs. (\ref{eq:10}),
\begin{equation}
\widehat{J}(\rho) =
\begin{pmatrix}
-1 +\partial \Psi/\partial \rho_e & \partial \Psi/\partial \rho_i \\
\alpha \partial \Psi/\partial \rho_e & -\alpha +\alpha \partial \Psi/\partial \rho_i
\end{pmatrix}
,
\label{Jacobian}
\end{equation}
calculated at the fixed points. The eigenvalues are
\begin{equation}
\lambda_{\pm}=-\frac{1}{2}(J_{11}+J_{22})\pm 
\frac{1}{2}\sqrt{(J_{11}-J_{22})^2+4J_{12}J_{21}},
\label{eigenv}
\end{equation}
where $J_{ij}$ are the entries of the Jacobian. If $\lambda_{\pm} <0$ at a fixed point, then this point is stable (attractor). If $\lambda_{\pm} > 0$, then the point is unstable.  If one of the eigenvalues $\lambda_{\pm}$ is positive and the other is negative, then the point is saddle. If Re$\lambda_{\pm} <0$ and Im$\lambda_{\pm} \neq 0$, the point is a stable spiral. If Re$\lambda_{\pm} > 0$ and Im$\lambda_{\pm} \neq 0$, the fixed point is an unstable spiral.
The real and imaginary parts of $\lambda_{\pm}$ determines the relaxation rate, $\gamma_r=-\text{Re}\lambda_{+}$, and the frequency, $\gamma_i=\text{Im}\lambda_{+}$, of damped oscillations.

\begin{table}[t]
\caption{
Stability of the fixed points 1, 2, and 3 in the regions I, II, and III on the phase diagram in Fig. \ref{fig-overview}, where s=stable, sd=saddle, u=unstable, sp=spiral, lc=limit cycle.}
\begin{center}
    \begin{tabular}{| c | c | c | c | c | c | c | c | c | c |}
    \hline
      & Ia   & Ib & Ic & Id      & Ie  & IIa  & IIb  & IIIa & IIIb   \\ \hline
    1 &  s   & s  & s  &  s      & s   &--    & --   & --   &  --      \\ \hline
    2 & --   & sd & sd & sd      & sd  &--    & --  & --   &  --      \\ \hline
    3 & --   & s  & s sp & u sp  & u &  s   & s sp & u sp \& lc & u \& lc \\ \hline
    \end{tabular}
\end{center}
\label{table1}
\end{table}

\begin{figure}
\includegraphics[width=0.4 \textwidth]{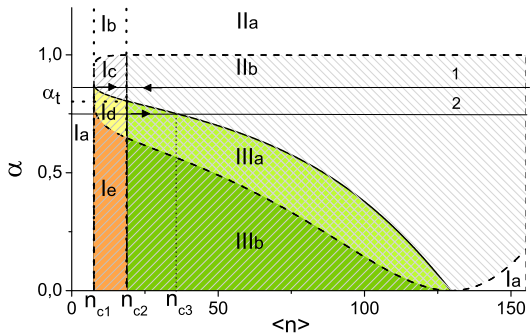}
\caption{(Color online) $\langle n \rangle -\alpha$ plane of the  phase diagram
of the cortical model.
$\langle n \rangle$ is the shot noise intensity. $\alpha$ is the ratio of the response time of excitatory neurons to the response time of inhibitory neurons. The phase regions,
the phase boundaries, and the parameters used in numerical calculations are explained in the text. $\alpha_t\approx 0.80$. Lines 1 and 2 represent two scenarios discussed in the text.
\label{fig-overview}}
\end{figure}

\emph{Phase diagram.} Analyzing the local stability of the fixed points
in the $\alpha-\langle n \rangle$ plane (see Table \ref{table1}), we found the phase diagram displayed in Fig.~\ref{fig-overview}.
In regions Ia-Ie, the network relaxes exponentially to the stable fixed point 1 (if perturbations are small). In regions Ib and IIa, relaxation to the stable fixed point 3 is exponential while, in regions Ic and IIb, the relaxation occurs in the form of damped oscillations. In regions IIIa and IIIb, the fixed point 3 is an unstable point surrounded by a limit cycle. These are the regions with sustained network oscillations about the point 3.
In Fig.~\ref{fig-overview}, the phase boundaries represented by the dashed and solid lines are determined by the conditions
$\gamma_i(\rho^{(3)})=0$ and $\gamma_r(\rho^{(3)})=0$,
respectively. Vertical lines represent lines $\langle n \rangle = n_{c1}$ and $n_{c2}$.

\begin{figure}
\includegraphics[width=0.45 \textwidth]{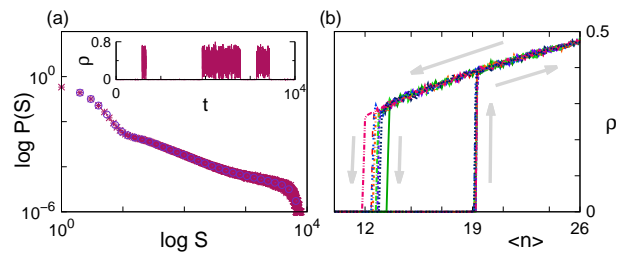}
\caption{(Color online)  (a)
Avalanche size distribution $P(s)$ versus size $s$ at $\langle n\rangle=18.8$. Inset: temporal activity of excitatory neurons
near the first-order phase transition. Time $t$ is in unit $1/\mu_e$. (b) Hysteresis in neuronal activity for increasing and decreasing noise level $\langle n\rangle$. In simulations, $\alpha=0.85$.
\label{fig-hysteresis}}
\end{figure}

\emph{First-order phase transition}.
The pattern of collective behavior depends on $\alpha$ and $\langle n\rangle$.
First we consider the case $\alpha >\alpha_t$
($\alpha_t$ is the critical value below which sustained oscillations appear).
In simulations and numerical solution of Eqs.~(\ref{eq:10}), we increased the noise level $\langle n \rangle$ from zero (region Ia) to a value in region IIa (or IIb) above the critical point $n_{c2}$ and afterwards decreased it again to a value below $n_{c2}$  (see line 1 in Fig.~\ref{fig-overview}).
Neuronal activity undergoes a jump at $\langle n \rangle = n_{c2}$
($ n_{c2} \approx 18.8$ in Fig.~\ref{fig-hysteresis}(b)).

\emph{Avalanches.} Near $n_{c2}$, we observed bursts of neuronal activity
caused by  avalanches (see the inset in Fig.~\ref{fig-hysteresis}(a)).
We studied avalanches,
analyzing spike time series by use of the standard method (see \cite{Beggs_2003} or the recent work \cite{Friedman_2012}).
The avalanche size distribution $P(s)$ is represented in Fig. \ref{fig-hysteresis}(a). When
$\langle n\rangle$ is close to
$n_{c2}$,  $P(s)$ is powerlaw, $P(s)\propto s^{-\sigma}$, in a broad range of $s$.
Using
the maximum likelihood estimate \citep{Touboul_2010}, we found $\sigma \approx 1.53$ and the corresponding p-value is $p=0.89$  (the closeness of $p$
to 1 shows that the fit is good).
This estimation agrees with the experimental data \cite{Beggs_2003,Friedman_2012} and the standard mean-field exponent $\sigma=3/2$ obtained for other exactly solved models \cite{Sethna_1993,Sethna_2001,Goltsev_2006}.
Thus, in the cortical model, bursts and avalanches are precursors of the first-order phase transition.
Another mechanism of avalanches based on self-organized criticality is discussed
in \cite{Chialvo_2006}.

\emph{Hysteresis.} If $\langle n \rangle$ decreases from a value
above $n_{c2}$ to a value below $n_{c2}$,
the network activity remaines as high as it was above $n_{c2}$ (see Fig. \ref{fig-hysteresis}(b)).
The activity falls to a low value only at $\langle n \rangle = n_{c1} < n_{c2}$. Hysteresis  occurs for $\alpha$ and $\langle n \rangle$ in regions Ib and Ic and is absent in regions Id and Ie where the fixed point 3 is unstable.
Hysteresis was observed, for example, in living neural networks \cite{Soriano} and in simulations of
thalamocortical systems \cite{izhikevich08}.

An analytical analysis of the cortical model gives us deeper understanding of the first-order phase transition. Solving the equation $\rho=\Psi(\rho,\rho)$
at $\langle n\rangle$ near $n_{c2}$ and substituting $\rho^{(1)}$ into Eq.~(\ref{eigenv}), we find that, in regions Ib and Ic
the relaxation rate to the low activity state $\rho^{(1)}$
is $\gamma_r\propto (n_{c2}-\langle n\rangle)^{1/2}$, i.e., $\gamma_r \rightarrow 0$
when $\langle n\rangle \rightarrow  n_{c2}$. This phenomenon is known as critical slowing down.
Solving Eqs.~(\ref{eq:10}) with weak white-noise forces $F_a(t)$  ($F_a(t) \propto 1/\sqrt{N}$), which mimic forces
caused by finite-size effects, we found, using the linear perturbation analysis \cite{Thomas_1982}, that the power spectral density (PSD) of the activity fluctuations, $S(\omega)=\langle \delta\rho_{a}(\omega) \delta\rho_{a}(-\omega)\rangle$, has a sharp zero frequency peak, $S(\omega)\propto 1/(\omega + \gamma_{r}^{2})$. If $\langle n\rangle \rightarrow  n_{c2}$, the peak maximum increases as $S_{max} \propto 1/\gamma_{r}^{2}$
(note that in the high activity state, there is no singularity around $n_{c2}$). If size $N$ is finite, then $\gamma_r$ remains nonzero, though very small, even at the critical point due to finite-size effects that smooth the transition observed in simulations. This leads to a finite value of $S_{max}$ at $n_{c2}$.

\emph{Non-equilibrium phase transitions}. An interesting scenario of collective dynamics takes place if inhibitory neurons respond slower than excitatory neurons and
$\alpha < \alpha_t$. In our analysis, we used simulations of the model, analytical and numerical solution of Eqs.~(\ref{eq:10}).
At a given $\alpha$, we start from $\langle n \rangle = 0$ and increase the shot noise intensity $\langle n \rangle$ (see line 2 in Fig.~\ref{fig-overview}). The neuronal network goes from region Ia with the single fixed point 1 into region Id or Ie where its dynamics
is determined by three fixed points (see Table \ref{table1}).
At $\langle n \rangle = n_{c2}$ the points 1 and 2 coalesce and the network undergoes a non-equilibrium phase transition due to the saddle-node bifurcation.
Above $n_{c2}$, the system is in region IIIa (or IIIb) with sustained oscillations (an unstable spiral or unstable fixed point surrounded by a limit cycle).
Then, at $\langle n \rangle =n_{c3}(\alpha) $ on the boundary between regions IIIa and IIb, the network undergoes a
phase transition due to the supercritical Hopf bifurcation. Above $n_{c3}$ the network enters region IIb with damped network oscillations
(a stable spiral in the point 3). The bifurcations were observed in the Morris-Lecar model stimulated by an applied current when the $I{-}V$ relation is N-shaped \cite{Rinzel1989}.
Thus, collective behavior of neuronal networks stimulated by a flow of random spikes bombarding neurons is similar to behavior
of a single neuron stimulated by an applied current. From this similarity it follows that the network acts as an `integrator', when it is close to the saddle-node bifurcation, and as a `resonator', when it is close to the Hopf bifurcation \cite{Izhikevich2000}.
The important difference between the models is that, in our case, we deal with
collective dynamics and phase transitions and it is necessary to find the order parameter and describe critical phenomena.


\begin{figure}
\includegraphics[width=0.45\textwidth]{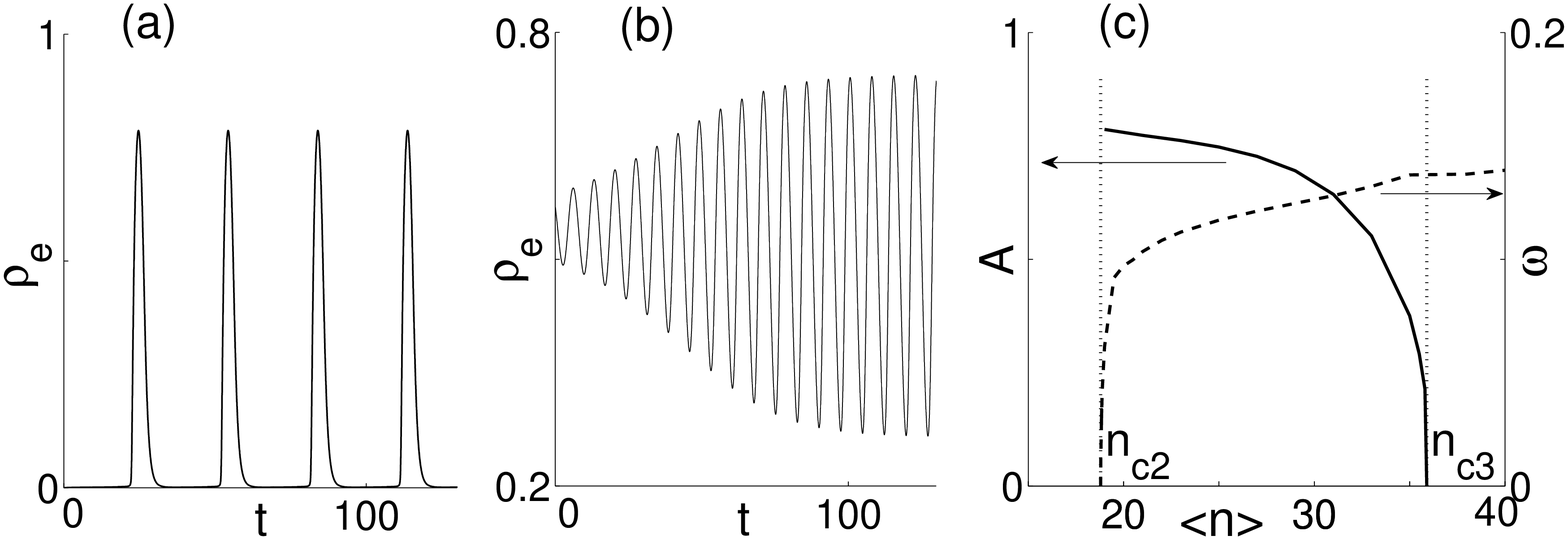}
\caption{
Network oscillations near (a) the saddle-node  ($\langle n \rangle=18.805, n_{c2}=18.8$) and (b) supercritical Hopf ($\langle n \rangle=34, n_{c3}=36$) bifurcations.
(c) Amplitude (solid line) and frequency (dashed line) of network oscillations versus
$\langle n \rangle$ (from a numerical solution of Eqs.~(\ref{eq:10})).
At $\langle n \rangle >n_{c3}$, the oscillations are damped.
Time $t$ is in units $1/\mu_e$, and $\alpha=0.75$.
\label{Hopf}}
\end{figure}

\begin{figure}
\includegraphics[width=0.45\textwidth]{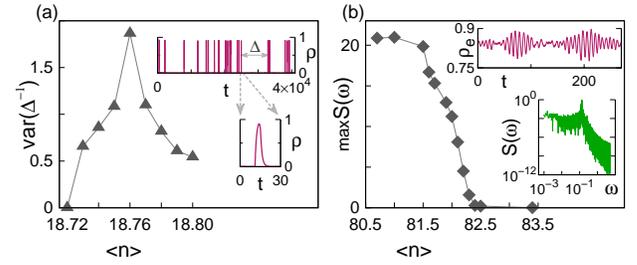}
\caption{(Color online)
(a) Variance of the reciprocal of interburst interval $\Delta$ versus $\langle n \rangle$ about the saddle-node bifurcation.
Inset: bursts of activity and a single burst at $\langle n \rangle=18.76$. Time $t$ is in units $1/\mu_e$.
(b) The peak maximum of the PSD
$S(\omega)$
of fluctuations versus $\langle n \rangle$ above the supercritical
Hopf bifurcation
($n_{c3}\approx 80.5$).
Inset: Temporal neuronal activity and $S(\omega)$  versus $\omega$ at $\langle n \rangle=82.5$.
In simulations, $\alpha=0.55$.
\label{precursors}}
\end{figure}

\emph{Saddle-node bifurcation}.
Above the saddle-node bifurcation, $\langle n\rangle >n_{c2}$, the network oscillations emerge with a large amplitude (see Fig.~\ref{Hopf}(a)) and their frequency increases from zero as $\omega \propto (\langle n \rangle - n_{c2})^{1/2}$  (see Fig.~\ref{Hopf}(c)). This behavior is generic for the bifurcation \cite{Strogatz_book1994,Rinzel1989,Izhikevich2000}.
We suggest that the frequency is the order parameter of the transition. In simulations, at $\langle n \rangle$ below $n_{c2}$, we observed  random almost identical
bursts of activity that are the manifestation of critical fluctuations
(inset of Fig. \ref{precursors}(a)). The bursts differ from ones found near the first-order phase transition in Fig. \ref{fig-hysteresis}(a). We assume that the bursts are non linear excitations generated by finite-size effects \cite{Explanation}.
The variance of the reciprocal of the interburst interval $\Delta $, $\langle [\Delta^{-1} -\langle \Delta^{-1} \rangle]^2\rangle$, where $\langle \Delta^{-1} \rangle$ is the mean burst frequency, has a peak at $n_{c2}$  (see Fig. \ref{precursors}(a)) and behaves as $|\langle n \rangle - n_{c2}|^{-1}$ about $n_{c2}$.
Burst properties and  generation  mechanisms need further investigations.

\emph{Supercritical Hopf bifurcation}.
Near the supercritical Hopf bifurcation at $ n_{c3}$, network oscillations differ from oscillations near the saddle-node bifurcation (see Figs.~\ref{Hopf}(a) and \ref{Hopf}(b)).
A decrease of the oscillation amplitude, $A \propto (n_{c3}  - \langle n \rangle)^{1/2}$,
and the relaxation rate, $\gamma_r \propto |n_{c3}  - \langle n \rangle|$, signals
the supercritical Hopf bifurcation (see Fig.~\ref{Hopf}(c)). This behavior is generic for the bifurcation \cite{Strogatz_book1994}. The amplitude
is the order parameter for the transition.
We found these results, expanding  $\Psi(\rho_e,\rho_i)$ in Eqs.~(\ref{eq:10}) in a series in $\delta\rho_{a}(t)\equiv\rho_{a}(t)-\rho^{(3)}$ around the fixed point 3 and holding terms up to $O(\delta\rho_{a}^{3})$ inclusively.
Then, we solved Eqs.~(\ref{eq:10}) in region IIIa,
using the averaging theory \cite{Strogatz_book1994}. If $\langle n\rangle$ tends to $n_{c3}$ from the `paramagnetic' region IIb with damped oscillations, then
fluctuations of neuronal activity are increased, signaling the bifurcation.
The PSD of the fluctuations
has a resonance peak at the frequency $\omega_0$ of damped oscillations: $S(\omega)/S_{max}\approx 4\zeta^2/[(1-x^2)^2+4\zeta^2 x^2]$,
where $x=\omega/\omega_0$, $\zeta=\gamma_r /\omega_0$ is the damping ratio, and the peak maximum $S_{max}\propto 1/\gamma_{r}^2$ (see also \cite{b2000,obh2009,lb2011}).
These results agree well with our simulations in Fig. \ref{precursors}(b).
They may help to understand the damping of alpha rhythms  by visual or auditory stimuli  \cite{Hari_1997,Lehtel_1997}.

In conclusion, using the exactly solvable cortical model of neuronal networks with stochastic neurons, we showed that a flow of random spikes bombarding neurons controls behavior of the network.
By increasing the flow, network oscillations appear at the saddle-node bifurcations and disappear at the supercritical Hopf bifurcation.
Burst, avalanches, and critical fluctuations of neuronal activity are precursors of the non-equilibrium transitions. Within the model, neuronal networks are excitable systems having dynamical behavior similar to the Morris-Lecar model of a biological neuron.

\begin{acknowledgements}
We thank S. N. Dorogovtsev and J. F. F. Mendes for stimulating discussions.
This work was  supported by the PTDC Projects No.
SAU-NEU/ 103904/2008, No. FIS/ 108476 /2008, No. MAT/ 114515 /2009, the project PEst-C / CTM / LA0025 / 2011, and the project "New Strategies Applied to Neuropathological Disorders," cofunded by QREN and EU.
K.~E.~L. and M.~A.~L. were supported by the FCT Grants No. SFRH/ BPD/ 71883/2010 and No. SFRH/ BD/ 68743 /2010.

\end{acknowledgements}



\end{document}